\documentclass[preprint,authoryear,12pt]{elsarticle}

\usepackage{graphics}
\usepackage{graphicx}
\usepackage{epsfig}
\usepackage{paralist}
\usepackage{gensymb}
\usepackage{xcolor,soul}
\usepackage{amssymb}

\journal{Advances in Space Research}

\begin{document}

\begin{frontmatter}

\title{A Roadmap towards a Space-based Radio Telescope for Ultra-Low Frequency Radio Astronomy
}

\author{M.J. Bentum\fnref{footnote2,footnote3}\corref{ca}}
\ead{m.j.bentum@tue.nl}

\author{M.K. Verma\fnref{footnote4}}
\ead{m.k.verma@student.tudelft.nl}

\author{R.T. Rajan\fnref{footnote4}}
\ead{r.t.rajan@tudelft.nl}

\author{A.J. Boonstra\fnref{footnote3}}
\ead{boonstra@astron.nl}

\author{C.J.M. Verhoeven\fnref{footnote4}}
\ead{C.J.M.Verhoeven@tudelft.nl}

\author{E.K.A. Gill\fnref{footnote9}}
\ead{e.k.a.gill@tudelft.nl}

\author{A.J. van der Veen\fnref{footnote4}}
\ead{a.j.vanderveen@tudelft.nl}

\author{H. Falcke\fnref{footnote5,footnote3}}
\ead{h.falcke@astro.ru.nl}

\author{M. Klein Wolt\fnref{footnote5}}
\ead{m.kleinwolt@astro.ru.nl}

\author{B. Monna\fnref{footnote6}}
\ead{b.monna@hyperiontechnologies.nl }

\author{S. Engelen\fnref{footnote6}}
\ead{s.engelen@hyperiontechnologies.nl}

\author{J. Rotteveel\fnref{footnote7}}
\ead{ j.rotteveel@isispace.nl}

\author{L.I. Gurvits\fnref{footnote8,footnote9}}
\ead{lgurvits@jive.eu}

\fntext[footnote2]{Eindhoven University of Technology, Faculty of Electrical Engineering,  5600 MB Eindhoven, The Netherlands}
\fntext[footnote4]{Delft University of Technology, Faculty EEMCS, Mekelweg 4, 2628 CD Delft, The Netherlands}
\fntext[footnote9]{Delft University of Technology, Faculty of Aerospace Engineering, Kluyverweg 1, 2629 HS Delft, The Netherlands}
\fntext[footnote3]{ASTRON, Netherlands Institute for Radio Astronomy, Oude Hoogeveensedijk 4, 7991 PD Dwingeloo, The Netherlands}
\fntext[footnote5]{Radboud University Nijmegen, P.O. Box 9010, 6500 GL Nijmegen, The Netherlands}
\fntext[footnote8]{Joint Institute for VLBI ERIC, Oude Hoogeveensedijk 4, 7991 PD Dwingeloo,  The Netherlands}
\fntext[footnote6]{Hyperion Technologies, Vlinderweg 2, 2623 AX Delft, The Netherlands}
\fntext[footnote7]{Innovative Solutions in Space, Motorenweg 23, 2623 CR Delft, Netherlands}

\begin{abstract}

The past two decades have witnessed a renewed interest in low frequency radio astronomy, with a particular focus on frequencies above 30~MHz e.g., LOFAR  (LOw Frequency ARray) in the Netherlands and its European extension ILT, the International LOFAR Telescope. However, at frequencies below 30~MHz, Earth-based observations are limited due to a combination of severe ionospheric distortions, almost full reflection of radio waves below 10~MHz, 
solar eruptions and the radio frequency interference (RFI) of human-made signals. Moreover, there are interesting scientific processes which naturally occur at these low frequencies. A space or Lunar-based ultra-low-frequency (also referred to as ultra-long-wavelength, ULW) radio array would suffer significantly less from these limitations and hence would open up the last, virtually unexplored frequency domain in the electromagnetic spectrum. 

A roadmap has been initiated by astronomers and researchers in the Netherlands to explore the opportunity of building a swarm of satellites to observe at the frequency band below 30~MHz. This roadmap dubbed Orbiting Low Frequency Antennas for Radio Astronomy (OLFAR), a space-based ultra-low frequency radio telescope that will explore the Universe's so-called dark ages, map the interstellar medium, and study planetary and solar bursts in the solar system and search them in other planetary systems. Such a radio astronomy system will comprise of a swarm of hundreds to thousands of satellites, working together as a single aperture synthesis instrument deployed sufficiently far away from Earth to avoid terrestrial RFI. The OLFAR telescope is a novel and complex system, requiring yet to be proven engineering solutions. Therefore, a number of key technologies are still required to be developed and proven. The first step in this roadmap is the NCLE (Netherlands China Low Frequency Explorer) experiment, which was launched in May 2018 on the Chinese Chang'e~4 mission. The NCLE payload consists of a three monopole antenna system for low frequency observations, from which the first data stream is expected in the second half of 2019, which will provide important feedback for future science and technology opportunities.

In this paper, the roadmap towards OLFAR, a brief overview of the science opportunities, and the technological and programmatic challenges of the mission are presented.

\end{abstract}


\end{frontmatter}

\parindent=0.5 cm

\section{Introduction}

One of the last unexplored domains of the electromagnetic spectrum unexplored by astronomers is the ultra-low frequency band. This frequency domain, which covers frequencies from $0.1$ to $30$~MHz, contains footprints of the early Universe (between $0.38$ and $400$ million years after the Big Bang) \citep{jester,rajan2016}, magnetospheric radio emissions from planets in the Solar System and perhaps from exoplanets \citep{zarka1998,zarka2007,Bentum2017a,Bentum2018}, and imprints of a broad variety of astrophysical processes \citep{jester}.

For a variety of reasons, astronomical observations in this frequency domain have been rare. One of the key reasons is a major impact of ionosphere on radio propagation. The ionosphere reflects and refracts radio waves at low frequencies, and completely blocks signals from space below the ionospheric cut-off frequency. The cutoff frequency depends on the time of the day and the activity of the Sun, but can be as high as $10$~MHz. In addition, human-made long range radio communications use this reflection property to broadcast signals across continents, as it keeps the radio waves inside the atmosphere (so called 'ducting effect'). This makes the situation with RFI for radio astronomy problematic, as it is almost impossible to find or create a radio quiet zone. Both these two factors make it impossible to make observations at the surface of the Earth in the ultra-low frequency domain. 

A space or Lunar-based radio telescope will overcome these problems and enable observations in the ultra-low frequency domain. There have been four space missions dedicated to low-frequency radio astronomical observations. The first radio astronomy explorer RAE-1 was deployed in an Earth orbit and made the first measurements of the Galaxy's spectrum, between $0.4$ and $6.5$~MHz \citep{alexander1969}. In 1971, the Interplanetary Monitoring Platform (IMP-6) conducted measurements of the Galactic radiation at 22 frequencies between 130 and 2600~kHz \citep{Brown1973}. The third mission was the radio astronomy explorer RAE-2 placed in a lunar orbit and observed between $25$~kHz and $13$~MHz \citep{alexander1975}. The fourth, and the most recent experiment on long-wavelength observations is the Netherlands Chinese Low Frequency Explorer (NCLE), which was launched as a part of the Chinese Chang'E~4 Lunar mission on May 21, 2018. Another synergistic development of the past decades is the implementation of the first generation Space Very Long Baseline Missions, the Japan-led VSOP/HALCA and Russia-led RadioAstron (\citet{Gurvits2019} and references therein).

The NCLE consists of three monopole antennas of 5 meter each, low noise amplifiers, a receiver system and a calibration unit, for monitoring the sky at low frequencies between $80$~kHz to $80$~MHz \citep{prinsloo2018}. In addition to these four implemented missions, there have been numerous concept studies for space-based radio arrays in the past \citep{french1967,weiler88,basart1997,jones2000}.
The more recent initiatives will be discussed in the next section. 

In order to match in angular resolution and sensitivity modern Earth-based low-frequency radio astronomy systems as LOFAR-ILT at ultra-low frequencies, a system with a synthesized aperture of at least $\sim100$~km is needed. Such a system can be composed of multiple satellites forming an interferometer, similar in principle to Earth-based low-frequency instruments e.g., LOFAR \citep{haarlem}. A space-based radio instrument as OLFAR will use small (nano-cube) satellites instead of conventional micro to large satellites. Several small satellites can be built with off-the-shelf components, making them affordable and simple to assembly. Hundreds to thousands of satellites can be deployed in a swarm formation, offering a good aperture ($u,v$) coverage and also baseline redundancy.

Satellite swarms in Earth orbits can in principle be formed while meeting the required aperture $(u,v)$-geometry and sufficient orbit  stability. However, such a system will not be protected against human-made RFI. In addition, the Auroral  Kilometre Radiation (AKR) \citep{Mutel2003} which is the strongest near-Earth source between a few tens of kHz and approximately 700~kHz, will contribute to the interference sustained by OLFAR. On Lunar orbits a swarm formation could be shielded from the terrestrial RFI \citep{BentumRFI2016}. However, recent studies have shown that on the Lunar orbits suitable for effective shielding, the relative motion of the satellites are unacceptably fast \citep{Bentum2019}.

A solution for reducing RFI might be deploying OLFAR at larger distances from Earth where the intensity of terrestrial RFI is lower. lower RFI allows for a smaller dynamic range of receiver systems on the satellites and therefore requires smaller data storage capacity. However, a larger distance over which the data need to be transmitted back to Earth increases the data transport challenge -- a clear subject of a trade-off consideration. Deploying OLFAR closer to Earth requires more data to achieve the same measurement fidelity compared with a higher altitude. On the other hand, the shorter distance to Earth makes it possible to transfer the data a a higher speed.

In this article the roadmap for implementing OLFAR is presented. Several intermediate steps are envisioned. The article is organized as follows. In Section~2, a brief overview of recent initiatives for an ultra-low frequency space-borne radio astronomy telescope is given. In Section 3, the OLFAR science cases and high level requirements are presented after which, in Section 4, the roadmap towards a full OLFAR is discussed. Conclusions are presented in Section 5.

\section{Overview of recent initiatives} 

During the last decade, several concept studies and projects have been drafted with the aim of building an ultra-low-frequency radio interferometer in space, and thus allows sky imaging down to a few kHz. This has long been impossible to implement due to the required aperture size. But it should be possible with current technology. The space-based studies have significantly improved the technology readiness levels (TRLs) of various subsystems, such as navigation, communications, small-spacecraft technology and swarm analysis.
At this moment several teams in the world are working towards an implementation of such a radio telescope, including:
\begin{itemize}
    \item In the Netherlands, a team consisting of several universities (Eindhoven University of Technology, Radboud University, Delft University of Technology), institutes (ASTRON, Joint Institute for VLBI ERIC, SRON), and companies (of which: Innovative Solutions in Space, Hyperion, NLR, TNO) started several initiatives: the ESA-funded DARIS (Distributed Aperture Array for Radio Astronomy In Space) project, \citep{DARIS, rajan2016}, the Dutch-funded OLFAR project \citep{Bentum2009,Rajan2011,rajan2016}, the ESA-proposed projects SURO (Space-based Ultra-long wavelength Radio Observatory) \citep{suro} and DSL (Discovering the Sky at the Longest Wavelengths, together with a Chinese team) \citep{boonstra2016} and the ESA-funded CLE (CubeSat Low Frequency Explorer) study  \citep{cle}. These studies made the foundation for the Dutch-Chinese NCLE (Netherlands-China Low Frequency Explorer) space-mission, which is a single-antenna satellite launched in 2018, located in the Moon-Earth L2 point, and which observes at low frequencies.
    \item In France, a NOIRE concept (Nanosatellites pour un Observatoire Interferometrique Radio dans l'Espace / Nanosatellites for a Radio Interferometer Observatory in Space) was proposed  by a team of several institutes led by the LESIA group \citep{noire2018}. The concept focused on the feasibility study of a swarm of nanosatellites equipped with  ultra-low frequency radio astronomy instruments. 
    \item A combined Swedish and UK team proposed the FIRST explorer \citep{first}, a constellation including six daughter spacecraft with radio astronomy antennas, and a mother spacecraft for science and metrology data processing and communications. 
     \item A number of initiatives and design studies of various concepts of ultra-low frequency missions have been presented over the past several decades in the USA. A comprehensive review of early studies is given by  \citet{Kassim-Weiler90}, \citet{basart1997} and references therein. The most recent USA initiative, RELIC \citep{RELIC} aims to deploy a multi-spacecraft constellation forming a decametric wavelength synthesized telescope on a selenocentric orbit. The science case of RELIC is presented extensively in about ten papers cited by \citep{RELIC}. The authors conclude that the proposed concept is viable for implementation on a rather short timescale of about five years. Another recent initiative in the USA, SunRISE \citep{SunRISE19} aims at studies of particle acceleration and transport in the inner heliosphere by means of radio interferometric imaging at frequencies from 100~kHz to 25~MHz. The concept assumes a formation of six satellites on a geosynchronous orbit with a typical baseline length between the satellites of about 10~km. The key technological idea behind the SunRISE concept is in utilizing commercial access to space, including a launch of all six spacecraft to their target orbit as a piggyback payload in conjunction with a larger spacecraft intended for GEO positioning.
     \item Several design studies of a broad range of prospective Space VLBI missions have been addressed in China over the past decade, including those aiming to operate in the ultra-low frequency domain \citep{An2019}. The NCLE experiment on the Chang'E-4 mission \citep{prinsloo2018} provided a new momentum for the ultra-low frequency studies which are envisioned in the framework of the DSL (Discovering the Sky at Longest wavelengths) initiative \citep{boonstra2016,Taikong14}.
   
\end{itemize}

In Table~\ref{tab:history}, an overview of the  initiatives is given, including the frequency range, maximum length of the baselines, the number of spacecrafts and the deployment location. All the initiatives, except NCLE, are project studies that have not proceeded to implementation stages. NCLE as well as OLFAR are discussed in more details in the following sections.

\begin{table}[]
    \centering
    \caption{History of space-based low frequency radio observatory studies.}
    \begin{tabular}{l|l|l|l|l|l} 
Name      &  Freq. range  & Baseline &  No. S/C  &  deploym. location & ref.  \\
          & [MHz]        & [km]         &           &                    & \\
\hline
DARIS  &  $1-30$      &    $\leq 100$      &  6-8      &    Dyn. Solar Orbit  & \citep{DARIS} \\
FIRST  &   $0.5-50$   &    $\leq 30$    &   7         &    Sun-Earth L2             & \citep{first} \\
SURO  &  $0.1–30$       &    $\leq 30$         &  8+        &     Sun-Earth L2              &  \citep{suro}  \\
DSL     &   $0.1-50$     &     $\leq 100$  &   8         &    Lunar Orbit          &  \citep{boonstra2016} \\
NOIRE &  $0.001-100$    &   $\leq 100$   &   various         &    Lunar Orbit          & \citep{noire2018} \\
RELIC     &    $0.1$-$30$  &  $\leq 100$  &   33         &    Lunar orbit  & \citep{RELIC} \\
NCLE   &    $0.08-80$     &     0        &   1       &   Earth-Moon L2             & \citep{prinsloo2018} \\
OLFAR &  $0.1-30$ &   100  &   $\geq 50$   &    TBD             & \citep{rajan2016}\\

         \hline
    \end{tabular}
    \label{tab:history}
\end{table}

\section{OLFAR -- Orbiting Low Frequency Antennas for Radio Astronomy }

Opening up the low frequency regime of 0.1- 30~MHz for radio astronomy requires a space-borne, in particular, Moon-based or Moon-orbiting, instrument as mentioned in the previous section. The objective of the OLFAR project is to design and build scalable autonomous satellite units to be used as an astronomical instrument at the ultra-low frequencies. In this frequency range, the sky brightness temperature can be as high at $10^7$~K at around $1$~MHz. The OLFAR instrument should be able to measure weak astronomical signals, orders of magnitude less bright than the sky itself, and will therefore need a very high dynamic range. In subsection \ref{requirememts} the requirements of the OLFAR instrument are presented. 

\subsection{Science Case} 

A number of science case studies or ultra-low frequency astronomy have been published over the past decades, e.g.  \citet{Gorgolewski1966,griessmeier2007predicting,jester,farside,boonstra2016,rajan2016,DARIS,noire2018, zarka2012} and references therein. We briefly summarize major topics of these cases, not attempting a comprehensive review of the above mentioned publications.

\subsubsection{Cosmology} 

The most popular high-profile science case for low frequency radio astronomy is the study of cosmology, especially  mapping the so called Dark Ages, \citet{jester} and references therein. It addresses the observation of the highly redshifted HI line, originated at the cosmological epoch before the formation of the first stars. Mapping and tomography of the HI distribution with adequate angular resolution, would require baselines of the order of tens of kilometers and a large number of antennas  \citep{jester}.

\subsubsection{Planetary radio emission} 

The magnetized planets in Solar System (Earth, Jupiter, Saturn, Uranus and Neptune) are emitting intense radio signals resulting from particle accelerations in various places of their magnetospheres. The auroral radio emission is emitted from above the magnetic poles of the planet through a cyclotron instability. The radiation belts (in the case of Jupiter and possibly the Earth) are emitting synchrotron radiation. The auroral radio emission, including AKR, carries imprints of the geophysical processes important for understanding the genesis and evolution of planets and their electromagnetic properties. Yet, comprehensive studies of AKR cannot be conducted from the Earth's surface due to the opaque ionosphere \citep{Mutel2003} thus requiring a space-borne facility.

The planetary atmospheric discharges can also be the source of radio pulses, which are associated with lightning. With an ultra-low-frequency space-based array, observations of lightning from planets in Solar system will become possible, including the Earth-originated lightning emission too. In addition, signals from magnetized exoplanets could also be detected. Given the prominence of the exoplanetary studies in the modern astronomy agenda, a contribution from ultra-low-frequency astronomy should be seen as strong science driver \citep{griessmeier2007predicting, zarka2012}.

\subsubsection{Galactic and stellar astrophysics} 

The low frequency sky mapping and monitoring of the galactic and extragalactic sky address a multitude of science topics on radio galaxies, active galactic nuclei, large scale structures like clusters with radio halos. As demonstrated by, e.g., a number of studies of galactic and extragalactic targets at decameter wavelengths reviewed by   \citet{Konovalenko+2016} and the ongoing LOFAR Two-Metre Sky Survey (LoTTS, \citet{LOTSS2017}), the decameter to meter wavelength domain offers a very reach scientific harvest. There are no reasons to believe that pushing the frequency domain to even lower frequencies will be less productive. Of special interest for studies at ultra-low frequencies are extremely redshifted galaxies as beacons in the early cosmological epochs. While a number of ground-based radio observatories have produced sky maps at frequencies close but still above the ionosphere cut-off at about 10~MHz, only very cursory sky mapping at frequencies below 10~MHz is currently available. The latter observations were provided by the Interplanetary Monitoring Platform (IMP-6) presented by \citet{Brown1973}, the the Radio Astronomy Explorers RAE-1 described by \citet{Alexander-Novaco-1974} and RAE-2 \citet{alexander1975,Novaco-Brown-1978}. The need for deeper sky mapping at ultra-low frequencies, especially in polarised emission is discussed by \citet{rajan2016} and references therein).

\subsubsection{Space Weather} 

An important gap in the current space weather research capabilities can be filled by a space-based array. A terrestrial low-frequency array such as LOFAR can observe in the frequencies as low as $30$~MHz. However, at lower frequencies a space-based instrument is needed, which will enable a wide range of solar and space weather research topics. These include for example, tracing the initial launch of a CME (coronal mass ejections), detailed tracking of the solar wind and CMEs through interplanetary space and in-depth studies of micro-structure in the Earth's ionosphere (\cite{oberoi2004lofar}).

\subsubsection{Ultra-high energy particles} 

Ultra-high energy particles are likely composed of  atomic nuclei, protons and potentially photons or neutrinos, which are extremely hard to detect due to their small flux of less than 1 particle per square kilometer per century, requiring large detectors such as the LOFAR. However, if a large array of low-frequency antennas are placed on the Lunar surface, then an increased sensitivity to cosmic rays and neutrinos can be exploited, as compared to the LOFAR \citep{Nelles_2015}. In addition, the radio quietness at the Lunar surface would greatly benefit such measurements \citep{scholten2006optimal}.

\subsection{OLFAR high level system requirements }
\label{requirememts} 

Table~\ref{tab:requirements} lists the high level system requirements of OLFAR derived from the science cases (see \citep{rajan2016}). The minimum distances between the satellites must be more than $10$~km and due to interstellar scattering the maximum baseline is limited to $100$~km, giving a resolution of $1$ arc-minute at $10$~MHz. OLFAR will comprise of more than $50$ satellites, each containing three linearly  polarized antennas, observing the sky from $0.1-30$~MHz. Extrapolation to higher frequencies will increase data bandwidth and processing overload. On the other hand, the number of satellites could vary from $50$ to larger than $10000$, and meet a diverse set of aforementioned science cases, as discussed in \citep{jester}. The satellites will employ passive formation flying and yet maintain sufficient position stability for a given integration time. In the presence of a stable orbit and thus stable baseline, position estimates can be more precisely known and thus the integration time can be extended up to $1000$ seconds and thereby reducing the down-link data rate \citep{rajan2016}.

\begin{table}[]
    \centering
    \caption{OLFAR system requirements}
    \begin{tabular}{l|l}
    \\
    \hline
Number of satellites  & $\geq$ 50 (scalable) \\
Number of antennas & $2$ or $3$ \\
Observation frequency range & $0.1 - 30$~MHz \\
Instantaneous bandwidth & $\geq$ 1~MHz \\
Spectral resolution & $1$~kHz \\
Snapshot integration time & $1$ to $1000$~s \\
Maximum baseline between satellites & 100~km \\
Deployment location & Moon orbit, Earth trailing, L2 \\
         \hline
    \end{tabular}
    
    \label{tab:requirements}
\end{table}

\section{Roadmap} \label{Roadmap}

\begin{figure} [h!]
\centering 
\includegraphics[width=1\textwidth]{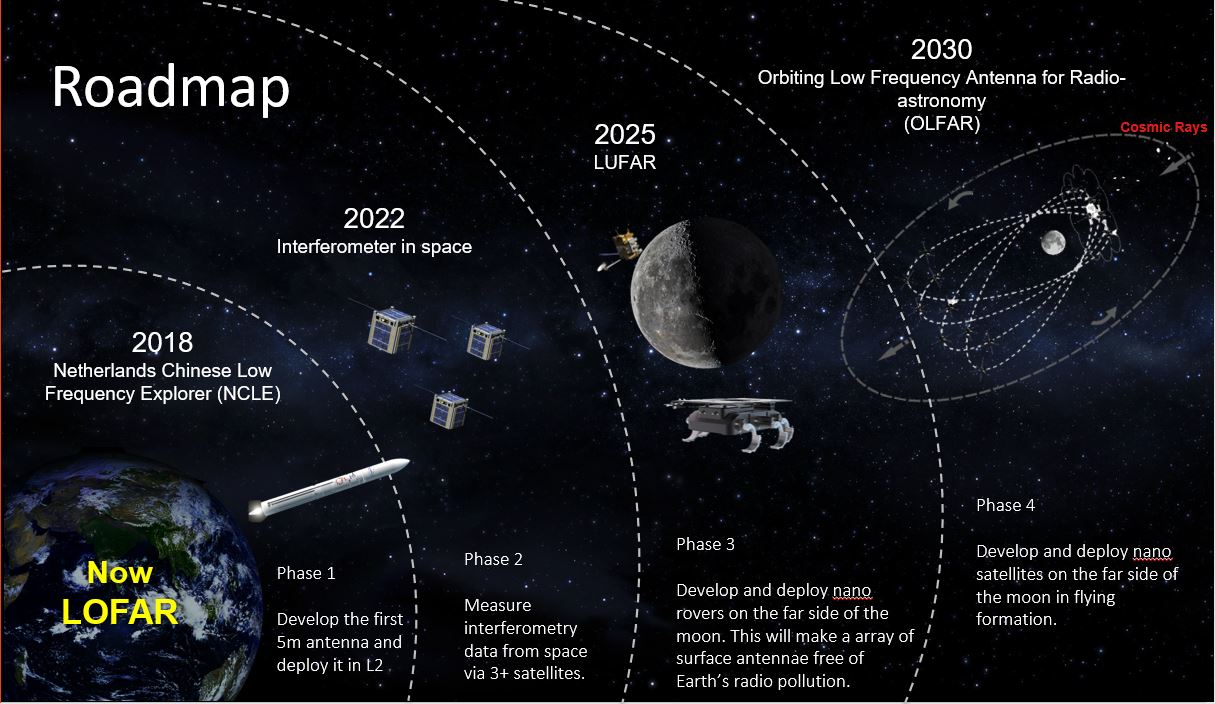} 
\caption{OLFAR Roadmap}
\label{OLFAR_Roadmap}
\end{figure}

A roadmap towards implementing the OLFAR mission is illustrated in Figure \ref{OLFAR_Roadmap}, which provides a wide range of technology candidates and a development pathway for the oncoming decade. The long term goal is to enable the deployment of OLFAR satellites in a Lunar orbit by 2030. Furthermore, the roadmap outlines the high-level requirements of each phase, and the necessary technological developments, which will be incorporated into the mission along the road to realizing OLFAR. 

The roadmap is divided into $4$ distinct phases, as illustrated in Figure \ref{OLFAR_Roadmap}, to gradually build up on operation experience, validate newly developed technologies, and address diverse science cases. Furthermore, the proposed roadmap is in alignment with the long term planning of other major space agencies, which allows for gradual build up of the swarm in the final location of OLFAR. 

In this section, we will outline all the four phases and provide a description, high level requirements, and the technologies that need to be developed to successfully realize the phase. We also mention the strategic importance of each phase in the context of the overall OLFAR mission.  To this end, each phase description is divided into key subsections, as follows. 
\begin{compactenum}
    \item Mission description
    \item Technology challenges
    \item High level requirements
    \item Time to launch
\end{compactenum}

\subsection{Phase 1: Netherlands Chinese Low Frequency Explorer (NCLE)} \subsubsection{Mission description:} The NCLE is a low-frequency payload which is currently on-board  the Chang'e 4 satellite (see Figure 2), which was launched on May 21th, 2018, and is located in a Lissajous orbit around the Earth-Moon L2 point (\cite{boonstra2017NCLE}). The NCLE design concept involves $3$ co-located, orthogonal, monopole antenna elements, each of $\sim$5~meters in length. These $3$ active antennas are mounted perpendicular to the upper side of the satellite body. The voltages between the monopoles and the spacecraft body are measured, digitized and send to Earth for further processing. The frequency range of NCLE is $80$~kHz to $80$~MHz, and overlaps with the Earth-based LOFAR LBA spectrum \citep{haarlem}, which could be useful for validation.

The NCLE is a pathfinder for a future space-based low-frequency radio interferometer, and thus one of the key objectives of NCLE is to  characterize the lunar radio environment and to understand the RFI at ultra-low frequencies. In addition, NCLE will also enable the study of solar activity, space weather, and possibly a detection of bright pulsars and other radio transient phenomena at ultra-long wavelengths, although the latter is not warranted due to the sensitivity limitations.

\begin{figure} [h!]
\centering 
\includegraphics[width=0.6\textwidth]{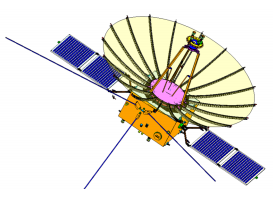} 
\label{Chang4}
\caption{An illustration of the Chang'e 4 satellite with three monopole antennas mounted on the body of the satellite (\cite{boonstra2017NCLE}).} 
\end{figure}

\subsubsection{Technological Challenges} One of the fundamental challenge in the NCLE payload is the design and development of the wideband science antenna and the following low-noise amplifier (LNA) (\cite{prinsloo2018}). The EMI modelling, system integration and calibration of the science front-end is a critical challenge in the NCLE mission. The analogue signals are digitized in a DSP system on which dedicated science modes are implemented in a flexible software-defined radio system, which is part of the NCLE payload. These modes for instance perform fast Fourier transforms to create average radio spectra, allow triggering on transient radio events, or allow to retrieve direction of arrival information using beam-forming or goniopolarimetry techniques (\cite{boonstra2017NCLE}). Finally, the data is relayed back Earth-based ground stations, by the Chang'e 4 satellite. The high level requirements of the NCLE Science instrument are listed in Table \ref{tab:Req_NCLE} \citep{prinsloo2018}.

\begin{table} [ht]
\label{tab:Req_NCLE}
\caption{High Level Requirements of the NCLE payload}
\centering
\begin{tabular}{l|p{0.8\linewidth}}
\hline
Req. ID & Requirement \\ \hline
NCLE1 & The NCLE antennas operates in the frequency range of 80~kHz to 80~MHz. \\
NCLE2 & Each of the three antennas is 5 m in length. \\
NCLE3 & NCLE instrument is adhere to electromagnetic interference specifications of MIL-STD-461.\\
NCLE4 & The location of the NCLE antennas on the spacecraft is not influencing the radiation performance of the antennas.\\
\hline
\end{tabular}
\end{table}

\subsection{Phase 2: LEO CubeSat Low Frequency Explorer (LCLE)} 

\subsubsection{Mission Description} The LCLE will be the first standalone mission to allow key \emph{interferometry technologies} to be tested for OLFAR. The mission will consist of 3 cubesatellites in a low Earth orbit (LEO), each equipped with 3 dipoles.  The low frequency measurements  will take place every time the satellites are at the apogee point of orbit, which is followed by correlation and integration of the measured data. The processed data is sent to the ground station for further analysis, when the satellites are nearest to Earth. 

\subsubsection{Technological Challenges} There are several technical advancements that are needed, to make this mission feasible. The LCLE is by no means the first mission to have cubesats flying as a group of three but it will be the first to deploy a radio sensitive payload in LEO and fly in a very tight formation.   The high accuracy needed to maintain flight within small position error means that the mission will heavily rely on an active attitude determination and control subsystem (ADCS) system. In the ADCS system, having a propulsion system which can provide enough delta V constantly maintaining the attitude and last for more than 3-4 years is one the biggest challenges to overcome for this mission. This also means a very accurate and reliable pointing algorithm is needed to make sure each satellite can independently locate its position and in relation to each other by sharing their own location details in the orbit. For inter-satellite communication, not only position data will be shared but also the raw measurements will be sent to each one of them to do distributed correlations. As each measurements will produce several Gigabits of data, a very high data inter-satellite communication is necessary which can be laser or RF based. The science front-end comprising of the antenna, antenna deployment mechanism, LNA and the digital signal processing system, will be miniaturized to enable power and mass efficiency. The high level requirements for the LCLE payload is listed in Table \ref{tab:Req_LCLE}.  

\begin{table} [ht]
\caption{High Level Requirements of LCLE}
\label{tab:Req_LCLE}
\centering
\begin{tabular}[width=\page]{l|p{0.8\linewidth}}
\hline
Req. ID & Requirement \\ \hline
LCLE1 & The science instrument shall operate in the radio frequency range of 1~MHz to 30~MHz  \\ 
LCLE2 & Inter satellite distance shall be 100~km at maximum.  \\ 
LCLE3 & Observations shall take place at least for a year.   \\ 
LCLE4 & Minimum inter-satellite communication data rate shall be 6 Mbps.   \\
\hline
\end{tabular}
\end{table}
          
\subsubsection{Estimated time to launch} The mission is expected to be ready for launch in 2022/2023. 

\subsection{Phase 3: LUnar low Frequency Antennas for Radio Astronomy (LUFAR)}
\subsubsection{Mission Description} LUFAR stands apart from the previously planned single-rover missions (see  \citep{farside}), as it makes use of several individual nodes, like LOFAR (\cite{haarlem}), to make a large array of antennas. Where all other concepts deploy large static stretched out antennas from a central node, a group of nanorovers (swarm) will carry one antenna each for LUFAR. State-of-the art rovers will enable a flexible spatial sampling of the UVW plane for radio astronomy observations. As the array is made up of mobile platforms, formation is dynamic in nature and can adjust in shape and size to test a number of configurations. 

The rovers will be deployed on the far side of the Moon onboard a lander which will land on the edge of the far side. This way, the lander can be in the field of view of the swarm on the far side and ground stations on Earth. This will allow the lander to act as a communication hub and relay data between the ground station, if a in-orbit relay spacecraft is not available. The rovers will take measurements and transfer raw data within the swarm for correlation via fibre optic cables. The processed data can be sent by the rovers to the relay system to be communicated to Earth. In the initial study, the number of rovers needed to make LUFAR successful for a wide range of science cases, is similar to OLFAR (50+ antenna systems). Therefore, the swarm will build itself gradually by adding 5 to 10 rovers each time a mission is flown to the Moon, if the ideal case  of a dedicated mission is not feasible.

Given results from the LCLE, the initial formation algorithm and on-board correlation, LUFAR will be implementing the same technology but at a much larger scale which will be a much closer representation of the OLFAR mission. One of the biggest advantages of testing high accuracy navigation subsystems and techniques on LUFAR is that the change in attitude is very low and the nodes are in much more stable and predictable relative position. This allows for much less risk factor and focus every aspect of the mission to optimize for the large number of nodes and mass production. Besides testing systems for lunar environment, an autonomous self-organization ('swarm-intelligence') algorithm will be implemented for the first time on a large scale and provide vital experience for the next phase of OLFAR. LUFAR will also validate the science cases possible that can be carried out on the far side of the Moon as rovers can maintain continuous observations for a long period of time, which is vital to know if OLFAR satellites are optimized  for the science cases described in Section 3.1.

LUFAR can be extended to allow the rovers to be used as a navigation reference for OLFAR satellites on lunar ground and provide auxiliary data for validation of observations made in orbit. In case of emergency, LUFAR may also provide a communications relay for flying nanosatellites and a data backup location. 

\subsubsection{Technological Challenges} LUFAR rovers will be deployed in harsh lunar environment, and carry out very sensitive measurements over a long period of time, in full autonomy. This mission therefore brings a relatively new list of technical challenges compared to the last two phases but is also critical to the next and final phase. 

For housekeeping and nominal operations of the rover, an autonomous swarming algorithm shall be developed which not only ensures the survival of the rover and swarm in very cold conditions but also navigates and maintains relative positioning. Antenna and its deployment system will also need to be developed to survive the abrasive and electrostatic nature of the dust particles. An 'intelligent' inter-rover communications will also be a demanding task to develop in order to prevent data congestion because a large amount of raw data will flow through this system. During night time, the rover should house a miniaturized very efficient power and active thermal system which will ensure critical system survive -180\degree C and have minimal electromagnetic interference with the measurements. One major bottleneck that is unique to LUFAR mission is the communication relay system. Although the lander and in-orbit satellite provide a means to do so, the data communications systems on board the rover needs to handle a very high data rate while also being miniaturized enough for a shoe-box sized rover, which is currently not available off-the-shelf. In addition, the LUFAR will establish inter-rover communication links for exchanging raw data, and enable distributed processing for correlation in the swarm and navigation. Furthermore, the processed data must be downlinked to Earth via an rover on the near-side or using a relay satellite. The high level requirements for the LUFAR mission are listed in Table \ref{tab:Req_LUFAR}.

\begin{table} [h]
\caption{High Level Requirements of LUFAR }
\label{tab:Req_LUFAR}
\centering
\begin{tabular}[width=\page]{l|p{0.8\linewidth}}
\hline
Req. ID & Requirement \\ \hline
LUFAR1 & The science instrument shall be able to operate in the radio frequency range of 0.1~MHz to~30~MHz.  \\ 
LUFAR2 & The distance between rovers shall be 10~km at maximum.  \\ 
LUFAR3 & Each rover shall have an internal operating temperature range of -40\degree\,C to 50\degree\,C.   \\ 
LUFAR4 & Each rover shall have the ability to be commanded from ground station individually.   \\ 
LUFAR5 & The science instrument's antenna shall be 10~m long. \\
LUFAR6 & Each rover shall not have a mass of more than 8~kg.   \\
LUFAR7 & The swarm shall be able to navigate and operate autonomously.  \\
\hline 
\end{tabular}
\end{table}
          
\subsubsection{Estimated time to launch} Delft university of technology and various Dutch partners are currently working on a concept of a Lunar Rover which will be a stepping stone for LUFAR. Given that NASA, ESA and other major agencies have similar plans and are up-scaling lunar missions in the coming decade, LUFAR has a good chance to be ready for flight by 2025, with full capability of swarm reached by 2030. 

\subsection{Phase 4: Orbiting low Frequency Antennas for Radio Astronomy (OLFAR)}
\subsubsection{Mission Description:} 
Our dot on the horizon, OLFAR, aims  to develop a space-based antenna array for radio astronomy comprising of 100's to 1000's of nano-satellites  for radio astronomy observations at 0.1--100~MHz. The satellites will  be spatially co-located within a maximum of 100~km, which will result in sky-maps of sub arc-minute resolutions at these wavelengths. This means that OLFAR will be 1-3 orders of magnitude more capable than its ground-based or space-borne predecessors in terms of sensitivity, dynamic range, and both angular and spectral resolution. The exact location of OLFAR is still under investigation. It could either be an orbit around the Moon or a location far far away from Earth to overcome the RFI from Earth. 

\subsubsection{Technological Challenges} 

To be able to launch OLFAR, a few technology breakthroughs will be needed. Many of the collaborators in aforementioned projects, including the NCLE, are part of our current consortium, offering a blend of expertise from radio astronomy, telecommunications, aerospace systems, embedded systems and navigation. A few of the fundamental science and technology challenges, which are listed in 5 distinct domains:

\begin{itemize}
    \item \emph{Array calibration and 3-D imaging} Antenna calibration is one of the biggest challenges for any low-frequency phased arrays, and for terrestrial  arrays, calibrating ionospheric phase fluctuations tops the list \citep{haarlem}. The instrumental beam shape, and the frequency dependent gains of the antennas must be calibrated on the fly with minimal Earth-assistance. For  low frequency observations, the requirements on positional stability and clock accuracies are in the order of sub-meters and sub-nanoseconds respectively, which drive the navigation requirements of OLFAR. Secondly, unlike terrestrial low-frequency arrays which are quasi-planar and have (at best) a hemispheric view, OLFAR will enable an instantaneous 3D view of the sky with a field-of-view of the full celestial sphere. 3D Imaging on a sphere with an interferometric array has been addressed by Carozzi \citep{Carozzi} as well as by Vugt et al. \citep{Vugt2017}.

    \item \emph{Navigation} The knowledge of position, time and orientation of individual satellites is vital for radio-science interferometry and for communication. However, due to the distant deployment and the shear number of satellites, current dependence on GPS-satellites and Earth-based ground segment are either not scalable or too expensive. In particular, when OLFAR is orbiting the far side of the Moon, the satellite swarm is completely disconnected from Earth for few hours \citep{Rajan2011}. This drawback calls for self-navigation or relative navigation of OLFAR, which suffices the needs of radio science, intra-satellite communication and collision avoidance. The desired synchronization accuracy is related to the accuracy of the onboard clocks and the observation frequencies of the OLFAR network \citep{rajan2013synchronization}. Despite satellite mobility, clock synchronization  can be achieved by employing generalized two-way ranging between mobile satellites \citep{rajanTSP2015}. Recent studies show that joint relative localization and synchronization is feasible for such GPS-denied or reference-free networks \citep{rajanESP2015,rajanESP2019}, however numerous questions still remain unanswered. For example, how can the relative time-varying position-time-orientation be estimated sufficiently accurate by a satellite swarm by only communicating with the neighbouring satellites? Can data fusion be applied to combine the satellite orbital dynamics and kinematics? Answering these questions will not only enable space-missions such as OLFAR, it would also contribute to breakthroughs in terrestrial navigation.
    \item \emph{Communications and Antenna Design} In the swarm of satellites there is a need for high-data-rate inter-satellite communication, both for localization and navigation purposes as well as for the exchange of data for the interferometric data processing. Multi satellite systems, operating as interferometers will generate so much data, that it needs to be processed to reduce the data volume to the essential information before it will be possible to relay the information onto the end user via a ground station. Apart from the need for high speed inter satellite communication, the need for high speed satellite-to-ground station communication is also rapidly growing. The total antenna system for the intended applications can be distinguished in three parts: the antenna system for the payload (the reception of astronomical signals), the antenna system for communication in the distributed space system and the antenna system for the data downlink to Earth.
    
    \item \emph{Space systems engineering and miniaturization} To establish a swarm of satellites of perhaps hundreds  of elements, miniaturization of the elements is required. This holds technology challenges on the level of individual elements, but also requires systems engineering breakthroughs in order to guarantee their efficient development, and a robust and efficient operations of the entire swarm and other mission elements. 
    
    Miniaturization may result in the design of small antennas. However, with smaller antennas, communication on lower frequencies is not possible. Higher frequencies are power hungry which might be a problem given that small harvesters provide limited energy. In the coming years, we want to address such problems both in terms of hardware and communication.
    
    Attitude control is necessary for antenna pointing and orientation and energy harvesters towards the energy source. To this end, we want to explore the ways to miniaturize attitude control systems such as momentum wheels, magneto-torquers, and our existing micro-propulsion system.

\item \emph{Distributed processing} The OLFAR swarm will employ distributed architectures for navigation, communication and processing science data, to avoid single point of failure, optimize efficiency and to abide by power constraints imposed by nano-satellites \citep{rajan2013} . While distributed computing in these individual domains are common in terrestrial applications, OLFAR  will juggle between these tasks while adhering to the strict science requirements levied on the mission.  OLFAR  will employ on-board distributed correlations to process streaming payload data, in order to reduce the data-downlink back to Earth. 

\end{itemize}

\section{Conclusion}
Space-based ultra-low-frequency radio astronomy will present a new milestone in our understanding of the Universe. It will open the frequency range below 30~MHz, which will enable unique science. For that purpose, missions such as OLFAR are essential to get access to this low-frequency part of the spectrum. 
The road to OLFAR, however, includes both scientific opportunities, and challenges concerning technological development and programmatics.
In this article, we briefly summarized the previous missions, and presented a roadmap to tackle the technological challenges for the dot on the horizon, a fully operational OLFAR.

\section{Acknowledgements}

We are grateful to the anonymous referees for very useful critique and constructive suggestions on the present paper.

\section{References}
\label{Section 3}

\bibliographystyle{apalike}
\bibliography{BibRoadmap}

\end{document}